\documentclass[prd,preprint,eqsecnum,showpacs]{revtex4}
\usepackage{amsmath}
\usepackage{graphicx}
\usepackage{bm}
\begin{document}
\title{Lorentz covariant statistical mechanics and thermodynamics of
the relativistic ideal gas and preferred frame}
\author{K. Kowalski, J. Rembieli\'nski and K.A. Smoli\'nski}
\affiliation{Department of Theoretical Physics, University
of \L\'od\'z, ul.\ Pomorska 149/153, 90-236 \L\'od\'z,
Poland}
\begin{abstract}
The Lorentz covariant classical and quantum statistical mechanics and 
thermodynamics of an ideal relativistic gas of bradyons (particles slower 
than light), luxons (particles moving with the speed of light) and tachyons 
(hypothetical particles faster than light) is discussed.  The Lorentz covariant 
formulation is based on the preferred frame approach which among others enables 
consistent, free of paradoxes description of tachyons.  The thermodynamic 
functions within the covariant approach are obtained both in classical and
quantum case.  
\end{abstract}
\maketitle
\section{Introduction}
Ideal gases are one of the most important model systems of the
nonrelativistic statistical mechanics and thermodynamics.  Examples
range from equations of state for classical gases to description
of electrons in metals and superconductors.  In spite of the fact
that the studies of a relativistic gas of massive particles
(bradyons, also called tardyons) go back to 1911 \cite{1}, the 
relativistic statistical mechanics and thermodynamics are far from 
complete.  On the one hand, the reason are apparent limited applications 
of the relativistic thermodynamics.  For example, in opinion of Ter Haar
and Wergeland \cite{2}:  ``At extremely high temperatures relativistic
effects may of course be important.  Then, however, matter behaves
as mixture of ideal gases and this limiting case poses no problem.
By and large, a relativistic theory of heat seems, therefore, to be
of little practical importance.''  Nevertheless, the arguments of
Ter Haar and Wergeland evidently fail for luxons and tachyons which
are relativistic particles regardless of the concrete value of the
temperature.  Furthermore, as pointed out by Arag\~ao de Carvalho
and Goulart Rosa \cite{3}, a fully relativistic treatment is required by
astrophysical systems such as white dwarfs and neutron stars.  On
the other hand, the development of the relativistic statistical
mechanics and thermodynamics was slowed down by the lack of the
covariant formulation.  In particular, we point out different transformation
rules of the relativistic temperature suggested by Einstein, Planck and von 
Laue, by Ott and by Landsberg \cite{2} .  The only formulation working in the case
of tachyons, based on the nonstandard (absolute) synchronization scheme, 
was introduced in a very recent paper \cite{4}. In this work we 
study the Lorentz covariant statistical mechanics and thermodynamics of the 
relativistic ideal gas of bradyons, luxons and tachyons.  In section 2
we recall the formulation of special relativity based on the absolute 
synchronization.  Section 3 is devoted to the classical
relativistic ideal gas.  In particular, we derive the covariant form of 
thermodynamic functions.  In section 4 we discuss the quantum statistical
mechanics and thermodynamics of the relativistic ideal gas.  Besides
derivation of the covariant forms of thermodynamic functions we also discuss 
the classical limit.
\section{Absolute synchronization scheme}
In this section we recall the basic facts about the formulation of special
relativity with the help of the absolute synchronization of clocks \cite{5}.  
Among others this approach provides a correct description of tachyons.  
Tachyons are hypothetical faster than light  particles. Besides their intriguing 
theoretically predicted properties \cite{6}, 
tachyons take attention of  physicists because they are candidates for the dark 
matter \cite{7} and dark energy \cite{8}.  Moreover they appear in brane theories 
such as the brane excitations  as well as in cosmological models (so called rolling 
tachyon models) \cite{9}. For this reason it is interesting to investigate 
statistical and thermodynamical properties of tachyonic gas. Unfortunately, 
standard description of tachyons is plagued by number of inconsistencies. Typical 
examples of such difficulties are the causal paradoxes (tachyon anti-telephone 
\cite{10}), the problem of so called transcendental tachyon (the space of tachyon 
velocities is not a Lorentz group carrier space \cite{11,5}) and vacuum instability 
on the quantum level \cite{12}.

As was stated many years ago by Sudarshan \cite{11} a consistent description 
of tachyons demands a preferred reference frame on the fundamental 
level. However, this means that the relativity principle is necessarily 
broken in such a case. This causes an apparent conflict with the standard 
Lorentz group transformations in the Minkowski space-time.  To overcome this 
difficulty let us notice that introduction of the inertial preferred frame 
(PF) means that we should realize the Lorentz group not only on the space-time 
coordinates but also on the four-velocity of the PF as seen by inertial observers. 
This gives us the necessary freedom to reconcile breaking of the relativity principle 
and simultaneously to preserve Lorentz covariance.  Such a realization of the Lorentz 
group was given in \cite{5} and it has an elegant explanation in terms of the bundle 
of frames as well as the physical interpretation in terms of the absolute 
synchronization scheme for clocks \cite{5,13,14}. In particular in \cite{5} a 
consistent classical and quantum description of tachyons was built in this framework, 
without of the above mentioned inconsistencies. It is important to stress that for 
massless and massive subluminal particles (luxons and bradyons respectively) this 
scheme is completely equivalent to the Einstein synchronization scheme (so called 
convention of synchronization \cite{5}) whereas it provides a consistent 
description of tachyons.  As an important application of the absolute 
synchronization in quantum mechanics the covariant relativistic position operator 
was recently introduced \cite{14}.  In the PF this operator reduces to the well-known 
Newton-Wigner operator. Finally, let us notice that the absolute synchronization is 
the most natural in cosmology because our universe distinguishes a frame (cosmic 
background radiation frame) as well as is flat on the large scales.  Let us also 
recall that some recent theoretical investigations of quantum gravity 
and extremely high energy phenomena predicts existence of a preferred frame. 
For example the loop gravity describing quantum gravitational phenomena at 
the Planck scale predicts the existence of a preferred frame and even incorporates 
quantum scale in the corresponding effective Lorentz group realization (the so 
called DSR theories \cite{15} and the Einstein-aether theories \cite{16}). 
Furthermore, in approach by Kosteleck\'y a breaking of Lorentz symmetry is assumed  
via specific field interactions \cite{17}.  Consequently, in this approach there 
exists effectively a preferred frame of reference too. 

Let us recall briefly the main results related to the description of 
tachyons in the framework of the absolute synchronization. As was stated above, 
in the absolute synchronization scheme the Lorentz transformations are realized 
simultaneously on the both coordinates in the inertial reference frames 
and velocity of PF, namely the contravariant transformation rules between
the frame $O_{u'}$ and $O_u$ are of the form
\begin{eqnarray}
x^{\prime}(u^{\prime})&=&D(\Lambda,u)x(u),\\
u^\prime&=&D(\Lambda,u)u.
\end{eqnarray}
where  $\Lambda$ is an element of the Lorentz group, $u$ is the four-velocity 
of the PF with respect to the inertial observer.  In (2.1) rotations are realized 
standardly i.e.\ $D(R,u)=R$, $R\in \rm{SO(3)}$, while boosts are $u$-dependent:
\begin{equation}
D(w,u)=\left(\begin{array}{c|c}
\frac{1}{w^0} & 0 \\
\hline
-\bm{w} & I + \frac{\bm{w}\otimes\bm{w}^T}%
{1+\sqrt{1+\bm{w}^2}} - u^0\bm{w}\otimes\bm{u}^T
\end{array}\right),
\end{equation}
where $w^\mu$ is four-velocity of the primed frame with respect to the 
unprimed one, and $\bm{w}\otimes\bm{u}^T$ designates the direct (Kronecker)
product of the column vector $\bm w$ and the row vector ${\bm u}^T$.  Hereafter 
the three-vector part of the covariant (contravariant) four-vector $a_\mu$ 
($a^\mu$) will be designated by $\underline{\bm a}$ ($\bm a$).  It is very 
important that the matrices $D(w ,u)$ are block-triangular, so the time 
coordinate is rescaled only by a positive factor. Namely,
\begin{equation}
x'^0 = \frac{1}{w_0}x^0.
\end{equation}
This enables us to avoid all inconsistencies  which plague the 
standard approach to tachyons \cite{6,10,11} because the notion of the 
instant-time hyperplane is invariant under Lorentz transformations. 
On the other hand, the form of (2.4) allows to solve problems arising in 
the standard approach, of covariance of some relativistic observables such as
the position operator mentioned earlier, related with mixing the time coordinate
with the spatial ones in the transformation rule for $x^0$.  From the technical
point of view the fact that in the standard Einstein synchronization the
time coordinate is mixed with the spatial ones is connected with the non-triangular
form of the Lorentz matrix given by (compare with the eq.\ (2.3))
\begin{equation}
\Lambda({\bm w}_{\rm E})=\left(\begin{array}{c|c}
w^0_{\rm E} & -{\bm w}^T_{\rm E} \\
\hline
-\bm{w}_{\rm E} & I + \frac{\bm{w}_{\rm E}\otimes\bm{w}^T_{\rm E}}%
{1+\sqrt{1+\bm{w}^2_{\rm E}}} 
\end{array}\right),
\end{equation}
where the subscript E designates the Einstein synchronization coordinates.
It is easy to see that the relationship between the coordinates in the Einstein
and the absolute synchronization is of the form
\begin{equation}
\begin{array}{ll}
x_{\rm E}^{0}=x^{0}+u^0{\bm u}\mbox{\boldmath${\cdot}$}{\bm x},
\qquad  & {\bm x}_{\rm E}={\bm x},\\\nonumber
u_{\rm E}^{0}=\frac{1}{u^0}, & {\bm u}_{\rm E}={\bm u}.
\end{array}
\end{equation}
Therefore the difference between both synchronizations lies in the definition of 
the time coordinate.  Notice that the time lapse in a fixed point ${\bm x}$ is the 
same in both synchronizations (i.e.\ if $d{\bm x}={\bm 0}$, then $dx^0_{\rm E}=dx^0$).  
Moreover, the transformation (2.1), similarly as the standard Lorentz transformation 
preserves the notion of an inertial frame.  The light velocity over a closed path is 
frame-independent as well.  Furthermore, from (2.6) it follows that both schemes 
(Einstein and absolute) are equivalent for velocities less or equal to the light
velocity (i.e.\ for bradyons and luxons), but for superluminal velocities (i.e.\ for 
tachyons) this equivalence is broken.  From the technical point of view such
nonequivalence is related to a singularity of the relationship between tachyon
velocities (derived from (2.6)) in both synchronizations.

The invariant Minkowski line element in the new coordinates in the frame $O_u$ has 
the following form:
\begin{equation}
ds^2=g_{\mu\nu}(u) dx^\mu dx^\nu,
\end{equation}
where the (covariant) metric tensor is frame-dependent (but not point-dependent!) 
and is given by
 \begin{equation}
 \left[g_{\mu\nu}(u)\right]=
 \left(\begin{array}{c|c}
 1 & u^0 \bm{u}^T \\
 \hline
 u^0 \bm{u} & -I + (u^0)^2 \bm{u}\otimes\bm{u}^T
 \end{array}\right).
 \end{equation}
The contravariant metric tensor is of the form
\begin{equation}
 \left[g^{\mu\nu}(u)\right]=
 \left(\begin{array}{c|c}
 (u^0)^2 & u^0 \bm{u}^T \\
 \hline
 u^0 \bm{u} & -I 
 \end{array}\right).
\end{equation}
Notice that from (2.9) it follows that the space metric is Euclidean in each
frame $O_u$ i.e.\ $dl^2 = d{\bm x}^2$ . Furthermore, $u^\mu u_\mu=1$, and 
space part of  the covariant four-velocity is equal to zero i.e.\ 
$\underline{\bm u} =\underline{\bm 0}$ in each frame, consequently  $u^0 =1/u_0$ . 
One can also easily check that we have
\begin{equation}
\frac{1}{(u^0)^2} - \bm{u}^2 = 1.
 \end{equation}
It is also useful to express the four-velocity $w$ in terms of the velocity 
${\bm v}={\bm w}/w_0$ of the primed frame with respect to the unprimed one 
by means of the formula
\begin{equation}
 w^0=\frac{1}{\sqrt{\left(1+u^0 \bm{u}\mbox{\boldmath${\cdot}$}\bm{v}\right)^2-
 \bm{v}^2}}
\end{equation}
obtained from the relation $w^\mu w_\mu=1$. 

The following remarks are in order.  The dispersion relations for four-momentum  
$p^\mu$ are given by
\begin{align}
p^\mu p_\mu &= m^2,&& \hbox{(bradyons)}\\
p^\mu p_\mu &= 0,&& \hbox{(luxons)}\\
p^\mu p_\mu &= -m^2.&& \hbox{(tachyons)}
\end{align}
We can solve these equations with respect to  $p^0$.  It follows that
\begin{align}
p^0 &= u^0\sqrt{({\bm u}{\bm\cdot}\underline{\bm p})^2+\underline{\bm
p}^2+m^2},&&\hbox{(bradyons)}\\
p^0 &= u^0\sqrt{({\bm u}{\bm\cdot}\underline{\bm p})^2+\underline{\bm
p}^2},&& \hbox{(luxons)}\\
p^0 &= u^0\sqrt{({\bm u}{\bm\cdot}\underline{\bm p})^2+\underline{\bm
p}^2-m^2}.&& \hbox{(tachyons)}
\end{align}
We point out that $p^0$ is in general different from the energy $p_0$ 
given by \cite{5}
\begin{align}
p_0 &= \frac{1}{u^0}[-{\bm u}{\bm \cdot}\underline{{\bm p}}+
\sqrt{({\bm u}{\bm\cdot}\underline{\bm p})^2+\underline{\bm
p}^2+m^2}],&&\hbox{(bradyons)}\\
p_0 &= \frac{1}{u^0}[-{\bm u}{\bm \cdot}\underline{{\bm p}}+
\sqrt{({\bm u}{\bm\cdot}\underline{\bm p})^2+\underline{\bm
p}^2}],&& \hbox{(luxons)}\\
p_0 &=\frac{1}{u^0}[-{\bm u}{\bm \cdot}\underline{{\bm p}}+
\sqrt{({\bm u}{\bm\cdot}\underline{\bm p})^2+\underline{\bm
p}^2-m^2}].&& \hbox{(tachyons)}
\end{align}
Now, the volume element $dV=d^3{\bm x}=dx^1\wedge dx^2\wedge dx^3$ 
transforms (under condition that $dx^0 =0$) according to 
\begin{equation}
dV' = w^0dV,
\end{equation}
so
\begin{equation}
V' = w^0V,
\end{equation}
where $V=\int_{t={\rm const}}dV$.  Similarly, taking into account that 
the Lorentz invariant momentum measure has the form 
\begin{equation}
d\mu(p) = \theta(p^0)\delta(p^2-m^2)d^4p,
\end{equation}
that is
\begin{equation}
d\mu(p) = \frac{d^3\underline{{\bm p}}}{2p^0},
\end{equation}
where $\underline{\bm p}$ is covariant momentum three vector, we deduce
that $d^3\underline{{\bm p}} = dp_1\wedge dp_2\wedge dp_3$  transforms 
as $d^3\underline {\bm p}'=(1/w^0)d^3\underline{\bm p}$.  Therefore we 
obtain the following formula on the Lorentz invariant phase space measure:
\begin{equation}
d\Gamma = -d^3{\bm x}d^3\underline{{\bm p}}.
\end{equation}
To complete the transformation rules discussed in this section we finally
write down the following relations:
\begin{equation}
{p^0}'=\frac{1}{w^0}p^0,\qquad {u^0}'=\frac{1}{w^0}u^0,\qquad {u_0}'=w^0u_0,
\end{equation}
which are also used in the next sections.
\section{Classical relativistic ideal gas}
We now discuss the basic properties of the classical ideal gas in
the absolute synchronization.  Taking into account the transformation properties 
discussed in the previous section it is easy to deduce that 
the Lorentz invariant partition function for each particle is given by \cite{4}
\begin{equation}
Z_1 = V\int d^3\underline{\bm p}\exp(-u_0^2\beta p^0),
\end{equation}
where $V$ is the volume of the system, $\beta=1/(kT)$ and temperature
transforms as $T'=w^0T$ under Lorentz transformations.  It should be
noted that in the preferred frame specified by $u_0=1$ and ${\bm u}={\bm 0}$,
$Z_1$ has the standard form.  Furthermore, by means of the eq.\ (3.1) and the 
standard definitions of the thermodynamical functions it is easy to show 
\cite{4} that temperature, internal energy, enthalpy, Helmholtz free energy 
and Gibbs free energy of the ideal gas transforms analogously to the volume 
(eq.(2.22)), whereas pressure, entropy and partition function are Lorentz 
invariant.  Consider the case of the relativistic ideal gas of $N$ 
noninteracting bradyons.  In the case of bradyons the partition function 
(3.1) takes the form
\begin{equation}
Z_{1b} = V\int d^3\underline{\bm p}\exp\left(-u_0\beta\sqrt{({\bm u}{\bm
\cdot}\underline{\bm p})^2+\underline{\bm p}^2+m^2}\,\right),
\end{equation}
where $m$ is the rest mass of a particle and we set $c=1$.  Using
the identities (A.1) and (A.2) we find
\begin{equation}
Z_{1b} = 4\pi V \frac{m^2}{u_0^2\beta}K_2(u_0\beta m),
\end{equation}
where $K_2(x)$ is the modified Bessel function (Macdonald function).
In the preferred frame, when $u_0=1$, we obtain from (3.3) the
well-known J\"uttner result \cite{1}.  Furthermore, the partition
function for luxons can be written as
\begin{equation}
Z_{1l} = V\int d^3\underline{\bm p}\exp\left(-u_0\beta\sqrt{({\bm u}{\bm
\cdot}\underline{\bm p})^2+\underline{\bm p}^2}\,\right),
\end{equation}
which leads with the use of (2.10) to
\begin{equation}
Z_{1l} = \frac{8\pi V}{u_0^4\beta^3}.
\end{equation}
Finally, in the case of tachyons the partition function is given by
\begin{equation}
Z_{1t} = V\int d^3\underline{\bm p}\exp\left(-u_0\beta\sqrt{({\bm u}{\bm
\cdot}\underline{\bm p})^2+\underline{\bm p}^2-m^2}\,\right).
\end{equation}
Taking into account (A.5) and (2.10) we get
\begin{equation}
Z_{1t} = 4\pi V \frac{m^2}{u_0^2\beta}S_{0,2}(u_0\beta m),
\end{equation}
where $S_{0,2}(x)$ is the Lommel function (see Appendix).  In the preferred 
frame when $u_0=1$, the formula (3.7) reduces to that originally obtained
by Mr{\'o}wczy{\'n}ski \cite{18}.

Now the partition function $Z$ for ideal gas of $N$ noninteracting
particles is
\begin{equation}
Z = \frac{1}{N!}Z_1^N.
\end{equation}
The knowledge of the partition function enables calculation of
thermodynamical quantities.  The average energy $U$ is related to
the partition function $Z$ by
\begin{equation}
U = -\frac{\partial}{\partial \beta}\ln Z.
\end{equation}
Hence, using (3.8), (3.3) and (A.2) we find the following formula on
the average energy of the bradyon ideal gas:
\begin{equation}
U_b = \frac{3N}{\beta} + N\frac{u_0mK_1(u_0\beta m)}{K_2(u_0\beta m)}.
\end{equation}
We remark that (3.9) implies the transformation rule for the average
energy of the form $U'=w^0U$.
From (3.10), (A.2) and (A.4) we find the following approximation of
$U_b$ for $u_0\beta m\gg 1$
\begin{equation}
U_b = \frac{3}{2}\frac{N}{\beta} + Nu_0m,\qquad u_0\beta m\gg 1.
\end{equation}
From (3.11) it follows that $U_b$ approaches $Nu_0m$ for large $\beta$
(see Fig.\ 1).  We point out that in the second extreme case 
$u_0\beta m\ll 1$, we have for an arbitrary thermodynamical quantity $X$
\begin{equation}
\lim_{(u_0\beta m)\to 0}X_{b,t}=X_l,
\end{equation}
following from the well-known fact that in the high-energy limit or
equivalently high-temperature limit bradyons and tachyons behave as
luxons.  

We now return to (3.9).  An immediate consequence of (3.8), (3.9)
and (3.5) is the following expression for the energy of the luxon
ideal gas:
\begin{equation}
U_l = \frac{3N}{\beta}.
\end{equation}
Notice that $U_l$ approaches zero as $\beta$ approaches infinity (see
Fig.\ 1).  Referring to (3.12) it is also clear that (3.13) is the limit of
(3.10) for $u_0\beta m\ll 1$.  Finally, taking into account (3.8),
(3.9), (3.7) and (A.7) we find that the average energy of the tachyon ideal
gas is given by
\begin{equation}
U_t = \frac{3N}{\beta} - N\frac{u_0mS_{-1,1}(u_0\beta
m)}{S_{0,2}(u_0\beta m)}.
\end{equation}
The relation (3.14) can be written in an equivalent form
\begin{equation}
U_t = -\frac{N}{\beta} + 3N\frac{u_0mS_{-1,3}(u_0\beta
m)}{S_{0,2}(u_0\beta m)}
\end{equation}
following directly from the second equation of (A.8).  In the
preferred frame, when $u_0=1$, (3.15) reduces to the formula
originally derived by Mr{\'o}wczy{\'n}ski \cite{18}.  The dependence
of the average energy of the ideal relativistic gas on the temperature
is shown in Fig.\ 1 and Fig.\ 2.  On using (3.14), (A.8) and (A.10) 
we arrive at the approximate relation:
\begin{equation}
U_t = \frac{2N}{\beta},\qquad u_0\beta m\gg 1.
\end{equation}
Therefore, as with luxons, for large $\beta$ the average energy $U_t$
approaches zero (see Fig.\ 1).
\begin{figure*}
\centering
\includegraphics[scale=.8]{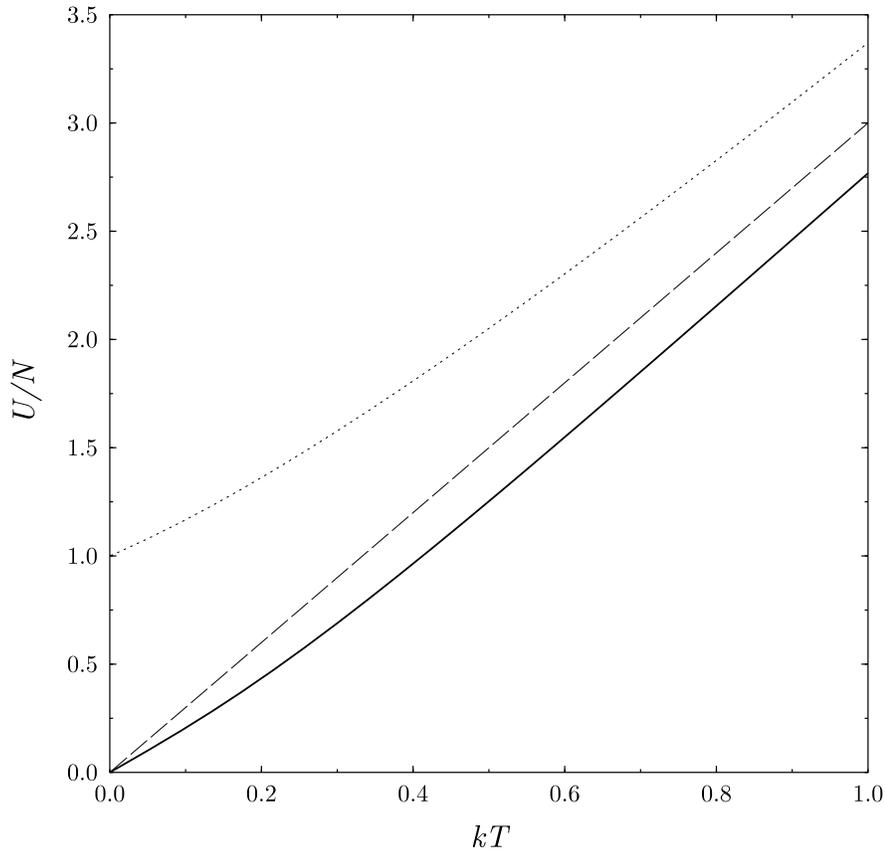}
\caption{The plot of the average energy $U$ per particle of the ideal 
gas of bradyons (dotted line), luxons (dash line) and tachyons 
(solid line) given by equations (3.10), (3.13) and (3.14), respectively, 
where $u_0=1$, and $m=1$ in (3.10) and (3.14), {\em versus\/} $kT$.
The behavior of average energy near the absolute zero.}
\end{figure*}
\begin{figure*}
\centering
\includegraphics[scale=.8]{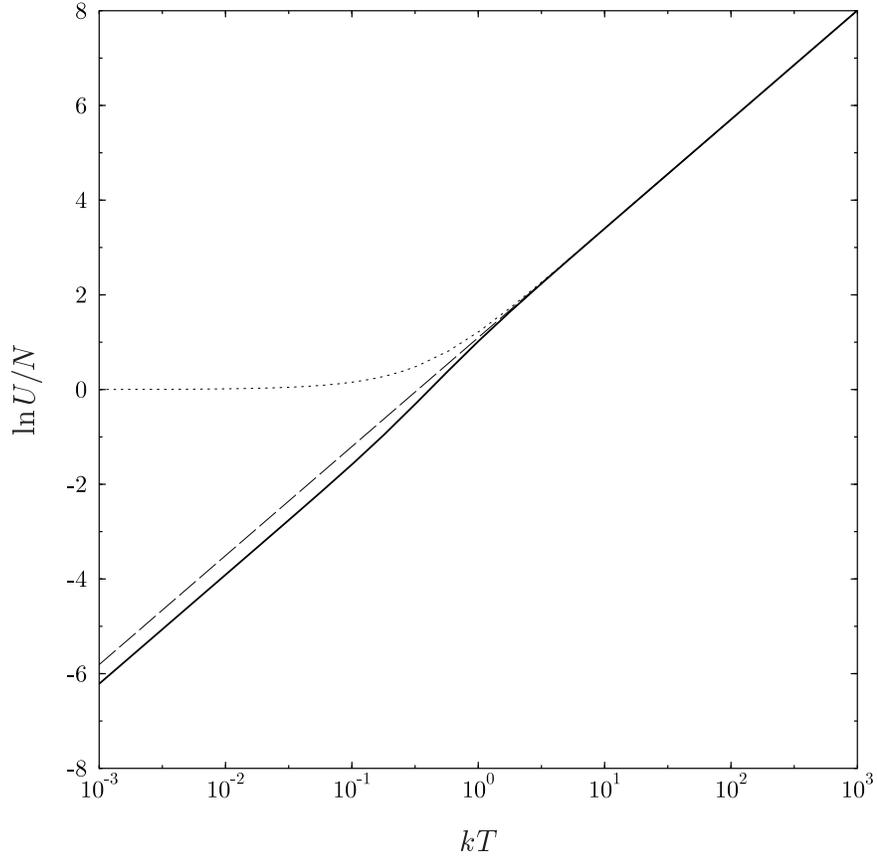}
\caption{Average energy $U$ per particle of the ideal gas of bradyons 
(dotted line), luxons (dash line) and tachyons (solid line) given by 
equations (3.10), (3.13) and (3.14), respectively, where $u_0=1$, and 
$m=1$ in (3.10) and (3.14), as a function of $kT$ on a logarithmic scale.  
The parameters $u_0$ and $m$ are the same as in Fig.\ 1.}
\end{figure*}

As an immediate application of the obtained relations (3.10), (3.13)
and (3.14) we now write down the formulas on the specific heat per
particle $C_v$ related to the average energy $U$ by
\begin{equation}
C_v = \frac{1}{N}\left(\frac{\partial U}{\partial T}\right)_V =
-\frac{k\beta^2}{N}\left(\frac{\partial U}{\partial \beta}\right)_V.
\end{equation}
Namely, we have for bradyons, luxons and tachyons, respectively
\begin{gather}
C_{vb} = k\left[3+(u_0\beta m)^2 -3\frac{u_0\beta mK_1(u_0\beta m)}
{K_2(u_0\beta m)}-\left(\frac{u_0\beta mK_1(u_0\beta
m)}{K_2(u_0\beta m)}\right)^2\right],\\
C_{vl} = 3k,\\
C_{vt} = k\left[3+\frac{u_0\beta mS_{-1,1}(u_0\beta m)-(u_0\beta
m)^2S_{-2,0}(u_0\beta m)}{S_{0,2}(u_0\beta m)}
-\left(\frac{u_0\beta mS_{-1,1}(u_0\beta
m)}{S_{0,2}(u_0\beta m)}\right)^2\right].
\end{gather}
The formula (3.20) can be written in an equivalent form
\begin{equation}
C_{vt} = -k\left[1+\frac{3[u_0\beta mS_{-1,3}(u_0\beta m)-5(u_0\beta
m)^2S_{-2,4}(u_0\beta m)]}{S_{0,2}(u_0\beta m)}
+\left(\frac{3u_0\beta mS_{-1,3}(u_0\beta
m)}{S_{0,2}(u_0\beta m)}\right)^2\right],
\end{equation}
following directly from (A.8).  Up to some typo in \cite{18} (3.21)
coincides in the preferred frame, when $u_0=1$, with the formula
originally obtained by Mr\'owczy\'nski.  We now return to (3.20). Taking 
into account (A.8) and (A.10) we obtain from (3.20) the following
asymptotic relation:
\begin{equation}
C_{vt} = 2k,\qquad u_0\beta m\gg 1.
\end{equation}
In the case of bradyons we find
\begin{equation}
C_{vb} = \frac{3}{2}k,\qquad u_0\beta m\gg 1.
\end{equation}
We remark that the asymptotic relations (3.22) and (3.23) can be formally
obtained from (3.17), (3.16) and (3.11).  Furthermore, it follows from 
numerical calculations that, in opposition to the case of bradyons when 
$C_{vb}$ is a decreasing function, the specific heat for tachyons has 
maximum (see Fig.\ 3).  The occurence of the maximum was treated by 
Mr\'owczy\'nski \cite{18} as a formal property of $C_{vt}$.  On the other hand, 
the maximum of the specific heat $C_v$ can be connected with the so called
Schottky anomaly in a two-state system \cite{19}.  In the preferred frame,
when $u_0=1$, the mass $m$ in the argument $u_0\beta m$ of the function
$C_{vt}$ is a counterpart of the energy $E$ of a two-state system exhibiting
Schottky anomaly, with a ground state of energy 0 and excited state of
energy $E$.  Although a physical interpretation of the maximum of the 
specific heat $C_{vt}$ is not clear, nevertheless, the experience with the
Schottky anomaly shows that the knowledge of the abscissa of the maximum 
of the specific heat would enable determination of the tachyon mass $m$.
\begin{figure*}
\centering
\includegraphics[scale=.8]{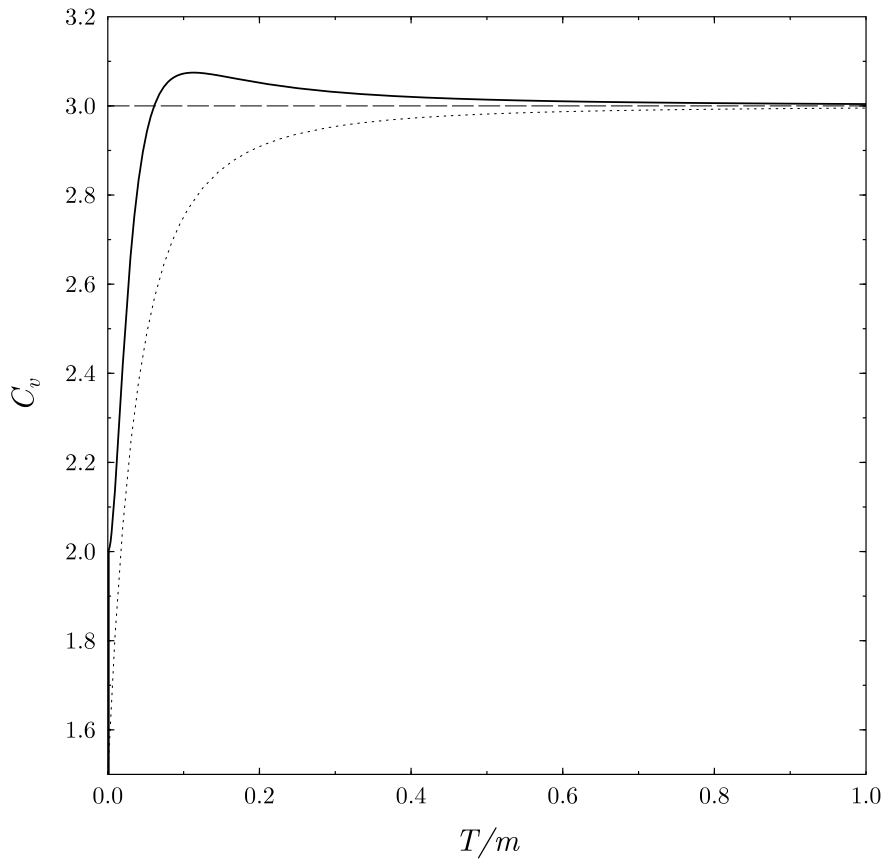}
\caption{The plot of the specific heat of the ideal gas of bradyons
(dotted line), luxons (dash line) and tachyons (solid line) given by 
(3.18), (3.19) and (3.20), respectively, where $u_0=1$ and $k=1$, 
{\em versus\/} $T/m$}
\end{figure*}

We now discuss the entropy of the relativistic ideal gas.  The Lorentz
invariant entropy $S$ can be expressed in terms of the partition function 
and the average energy.  Namely,
\begin{equation}
S = k\ln Z + \frac{U}{T}.
\end{equation}
Using (3.24), (3.8) and the Stirling formula
\begin{equation}
N! \approx \left(\frac{N}{e}\right)^N,
\end{equation}
which holds for large $N$, we get
\begin{equation}
S = kN\ln Z_1 -N\ln N + N + \frac{U}{T}.
\end{equation}
From (3.26), (3.3) and (3.10) we obtain in the case of bradyons
\begin{equation}
S_b = kN\left(\ln\frac{4\pi Vm^3}{u_0N}+\ln\frac{K_2(u_0\beta
m)}{u_0\beta m} + \frac{u_0\beta mK_1(u_0\beta m)}{K_2(u_0\beta m)}
+ 4\right).
\end{equation}
The formula on the entropy for the ideal gas of luxons such that
\begin{equation}
S_l = kN\left(\ln\frac{8\pi V}{u_0^4N\beta^3} + 4\right)
\end{equation}
is implied by (3.26), (3.5) and (3.13).  Finally, using (3.26),
(3.7) and (3.14) we find that the entropy of the ideal gas of
tachyons can be written as
\begin{equation}
S_t = kN\left(\ln\frac{4\pi Vm^3}{u_0N}+\ln\frac{S_{0,2}(u_0\beta
m)}{u_0\beta m} - \frac{u_0\beta mS_{-1,1}(u_0\beta m)}{S_{0,2}(u_0\beta m)}
+ 4\right).
\end{equation}
The dependence of the entropy of a relativistic ideal gas on the 
temperature is shown in Fig.\ 4.  We remark that, as in the nonrelativistic
case, the entropies (3.27), (3.28) and (3.29) approach minus infinity
at the absolute zero.  We recall that such behavior of the entropy means
that the classical physics fails in the limit of very low temperatures
and one should employ quantum statistical mechanics to calculate the entropy.
\begin{figure*}
\centering
\includegraphics[scale=.8]{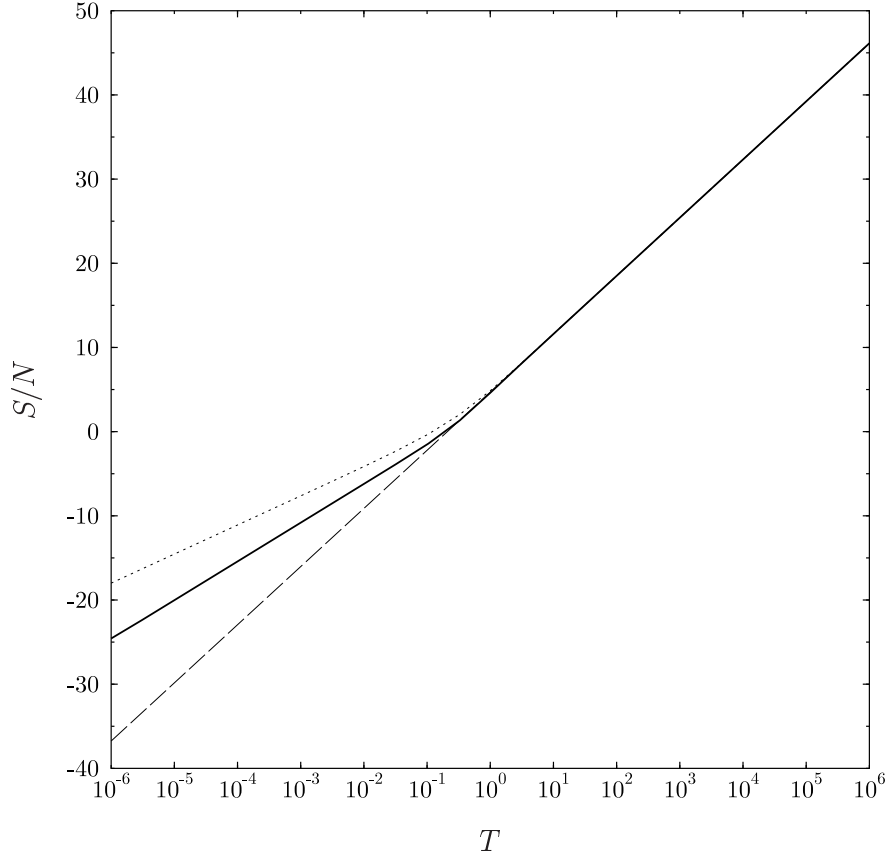}
\caption{A semilogarithmic graph in which entropy per particle of the 
ideal gas of bradyons (dotted line), luxons (dash line) and tachyons 
(solid line) given by equations (3.27), (3.28) and (3.29), respectively, 
where $u_0=1$, $k=1$ and $m=1$, is plotted against the temperature.}
\end{figure*}

Our purpose now is to study the Helmholtz free energy.  The free
energy $F$ is related to the partition function $Z$ by
\begin{equation}
F = -\frac{1}{\beta}\ln Z.
\end{equation}
Using (3.30), (3.3), (3.5), (3.7) and the Stirling formula we obtain
the following formulas on the free energy for bradyons, luxons and
tachyons, respectively:
\begin{eqnarray}
F_b &=& -\frac{N}{\beta}\left(\ln\frac{4\pi Vm^3}{u_0N}+\ln\frac{K_2(u_0\beta
m)}{u_0\beta m} + 1\right),\\
F_l &=& -\frac{N}{\beta}\left(\ln\frac{8\pi V}{u_0^4N\beta^3} +
1\right),\\
F_t &=& -\frac{N}{\beta }\left(\ln\frac{4\pi Vm^3}{u_0N}+\ln\frac{S_{0,2}(u_0\beta
m)}{u_0\beta m} + 1\right).
\end{eqnarray}
As an immediate consequence of the above relations, and the formula on
the chemical potential such that
\begin{equation}
\mu = \left(\frac{\partial F}{\partial N}\right)_{T,V}.
\end{equation}
we get
\begin{eqnarray}
\mu_b &=& -\frac{1}{\beta}\left(\ln\frac{4\pi Vm^3}{u_0N}+\ln\frac{K_2(u_0\beta
m)}{u_0\beta m}\right),\\
\mu_l &=& -\frac{1}{\beta}\ln\frac{8\pi V}{u_0^4N\beta^3},\\
\mu_t &=& -\frac{1}{\beta }\left(\ln\frac{4\pi Vm^3}{u_0N}+\ln\frac{S_{0,2}(u_0\beta
m)}{u_0\beta m}\right).
\end{eqnarray}
Now, making use of the formula on the pressure $p$ such that
\begin{equation}
p = -\left(\frac{\partial F}{\partial V}\right)_T,
\end{equation}
and the above formulas on the free energy we arrive at the equation
of state which is the same for bradyons, luxons and tachyons, namely
\begin{equation}
pV = kTN.
\end{equation}
Bearing in mind the formula (3.39) one may conclude, as for example
Mr\'owczy\'nski \cite{18}, that ``{\ldots} all properties of the
classical gas of tachyons and the gas of bradyons are similar and no
new phenomena have been found in the case of tachyons''.  We do not
share such opinion.  First of all it seems that the better version
of the equation of state than (3.39) which does not distinguish
bradyons, luxons and tachyons is the equation of state envolving the
density of energy $U/V$ instead of the density of particles $N/V$
used in (3.39).  Indeed, using (3.39), (3.10), (3.13) and (3.14) we
get
\begin{equation}
p = \frac{\rho}{f_{b,l,t}(u_0\beta m)},
\end{equation}
where $\rho=U/V$ is the density of energy and the functions $f_b$,
$f_l$ and $f_t$ corresponding to the case of bradyons, luxons and
tachyons, respectively, are given by
\begin{eqnarray}
f_b(u_0\beta m) &=& 3+\frac{u_0\beta mK_1(u_0\beta m)}{K_2(u_0\beta
m)},\\
f_l(u_0\beta m) &=& 3,\\
f_t(u_0\beta m) &=& 3-\frac{u_0\beta mS_{-1,1}(u_0\beta m)}{S_{0,2}
(u_0\beta m)}.
\end{eqnarray}
The plot of the above functions is given in Fig.\ 5.  We point out that
in the high-temperature limit we have (see (3.12))
\begin{equation}
\lim_{(u_0\beta m)\to 0}f_{b,t}(u_0\beta m)=f_l(u_0\beta m)=3,
\end{equation}
and in the limit $T\to 0$
\begin{equation}
f_b(u_0\beta m)=u_0\beta m,\qquad u_0\beta m\gg 1,
\end{equation}
and
\begin{equation}
f_t(u_0\beta m)=2,\qquad u_0\beta m\gg 1,
\end{equation}
following directly from (3.11) and (3.16), respectively.
\begin{figure*}
\centering
\includegraphics[scale=.8]{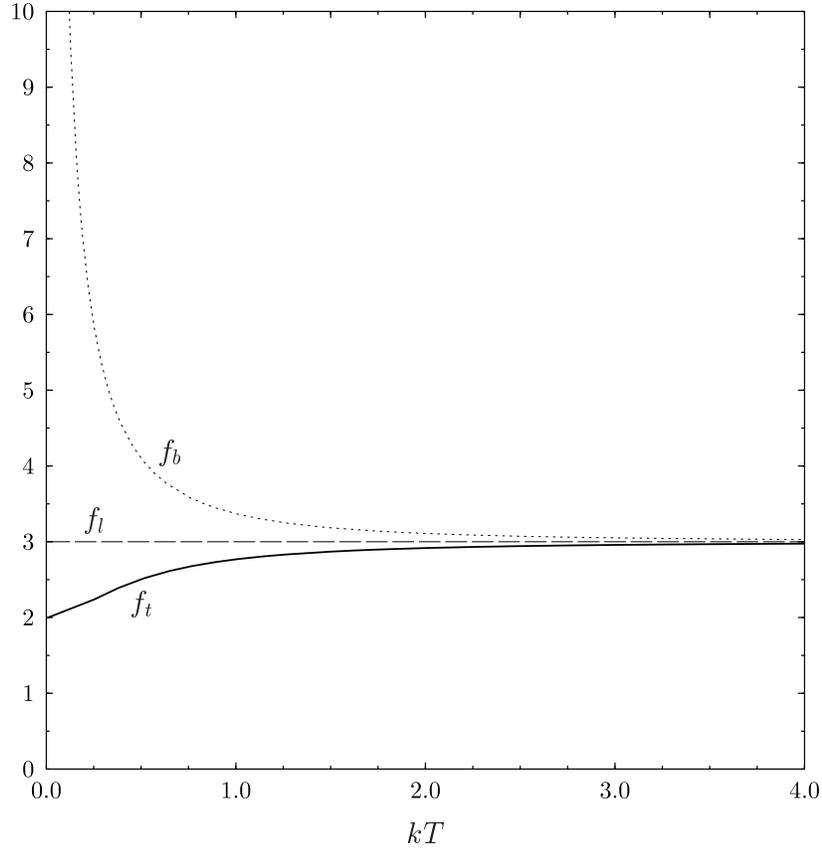}
\caption{The plot of the function $f_b$ (dotted line), $f_l$ 
(dash line) and $f_t$ (solid line) given by (3.37), (3.38) and (3.39), 
related to the equation of state (3.36) for bradyons, luxons and 
tachyons, respectively, where $u_0=1$, and $m=1$, in (3.37) and
(3.39), {\em versus\/} $kT$.}
\end{figure*}
\begin{figure*}
\centering
\includegraphics[scale=.8]{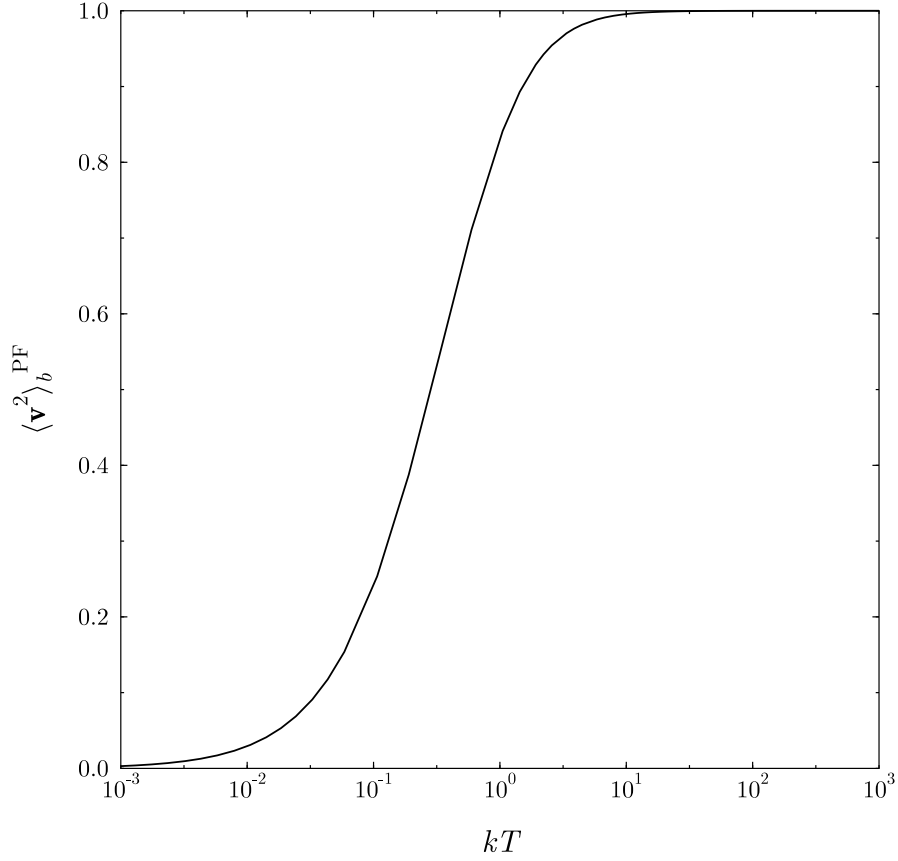}
\caption{A semilogarithmic graph in which squared velocity of particles of 
the ideal gas of bradyons given by (3.47), where $m=1$, is plotted against
$kT$ on a logarithmic scale.}
\end{figure*}
\begin{figure*}
\centering
\includegraphics[scale=.8]{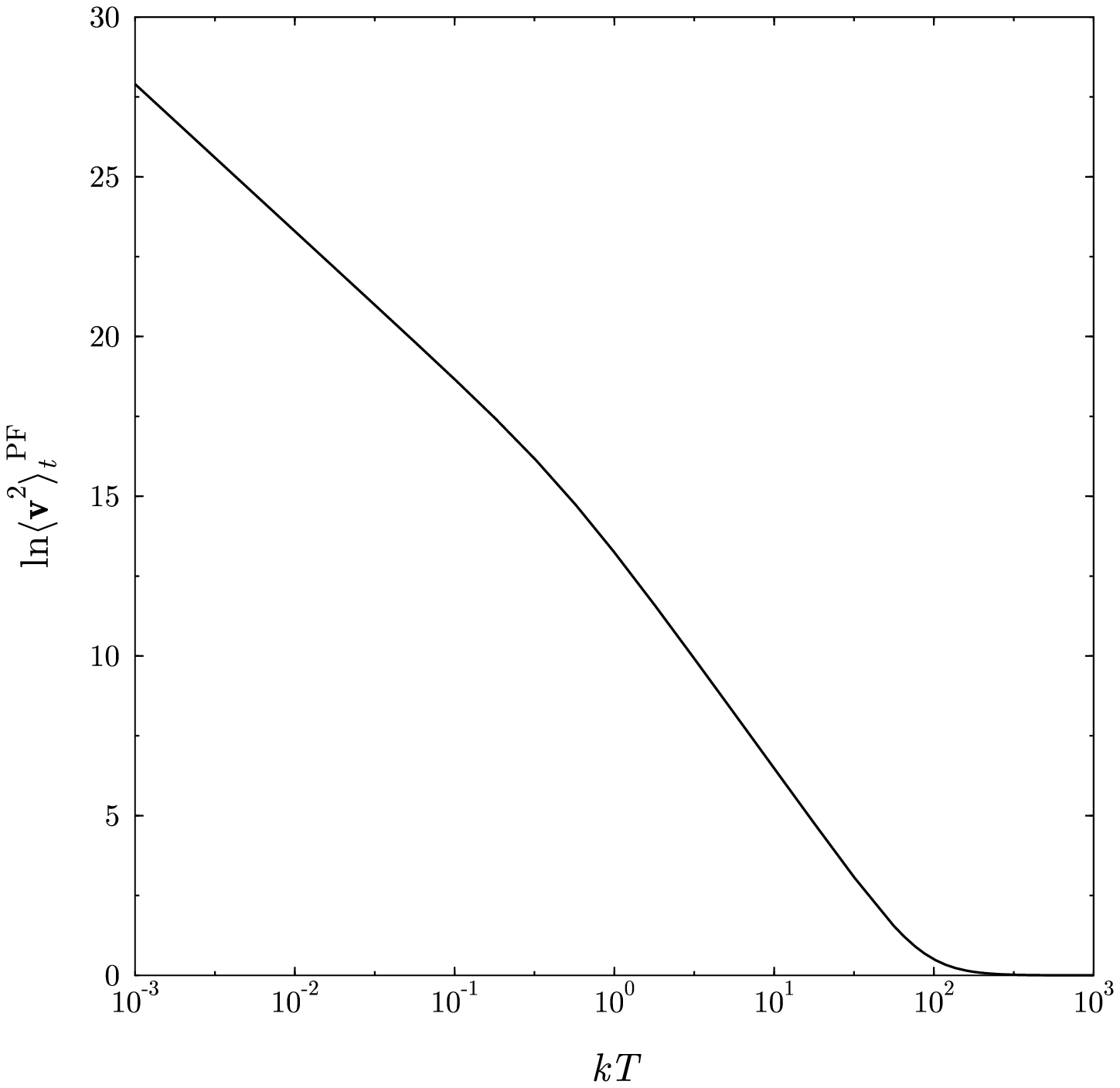}
\caption{Average squared velocity of particles of the ideal gas of tachyons 
as a function of $kT$ (see 3.48) on a logarithmic scale. The parameter $m$
is the same as in Fig.\ 6.}
\end{figure*}
Furthermore, in opposition to Mr\'owczy\'nski, we treat the occurrence of 
the maximum of specific heat for tachyons as their important property 
distinguishing them from both bradyons and luxons.  Even more evident 
difference of behavior of tachyons and bradyons is the dependence of velocity on
the temperature.  Indeed, it is clear in view of the fact that 
upon losing energy tachyon accelerate, that the velocity of a tachyon 
should be decreasing function of temperature.  Of
course, it is not the case for bradyons.  To be more specific,
consider the average squared velocity $\langle {\bm v}^2\rangle$.
For the sake of simplicity we restrict to the preferred frame, i.e.\
we set $u_0=1$ and ${\bm u}={\bm 0}$.  The average squared velocity
for bradyons is given by
\begin{multline}
\langle {\bm v}^2\rangle_b^{\rm PF} = \frac{V\int d^3\underline{\bm p}\,
{\bm v}^2\exp\left(-\beta\sqrt{\underline{\bm p}^2+m^2}\right)}{Z_{1b}^{\rm PF}} =
\frac{V\int d^3\underline{\bm p}\,\frac{\underline{\bm
p}^2}{\underline{\bm p}^2+m^2}
\exp\left(-\beta\sqrt{\underline{\bm p}^2+m^2}\right)}{Z_{1b}^{\rm PF}}\\
= \frac{m\beta\int_1^\infty x^{-1}(x^2-1)^\frac{3}{2}
e^{-m\beta x}dx}{K_2(m\beta)},
\end{multline}
where $Z_{1b}^{\rm PF}=Z_{1b\vert_{u_0=1}}$ (see (3.3)).  The plot of 
$\langle {\bm v}^2\rangle_b^{\rm PF}$ {\em versus\/} $\beta$ is
shown in Fig.\ 6.  The average squared velocity for tachyons in the
preferred frame can be written as (see Fig.\ 7)
\begin{multline}
\langle {\bm v}^2\rangle_t^{\rm PF} = \frac{V\int d^3\underline{\bm p}\,
{\bm v}^2\exp\left(-\beta\sqrt{\underline{\bm p}^2-m^2}\right)}{Z_{1t}^{\rm PF}} =
\frac{V\int d^3\underline{\bm p}\,\frac{\underline{\bm
p}^2}{\underline{\bm p}^2-m^2}
\exp\left(-\beta\sqrt{\underline{\bm p}^2-m^2}\right)}{Z_{1t}^{\rm PF}}\\
= \frac{m\beta\int_0^\infty x^{-1}(x^2+1)^\frac{3}{2}
e^{-m\beta x}dx}{S_{0,2}(m\beta)},
\end{multline}
where $Z_{1t}^{\rm PF}=Z_{1t\vert_{u_0=1}}$ (see (3.7)).
\section{Quantum relativistic ideal gas}
Our purpose now is to study the quantum ideal gas in the absolute
synchronization.  The classical formula (3.1) and the nonrelativistic
relations in the case of the canonical ensamble \cite{20} indicate the
following form of the partition function for the relativistic gas of
$N$ particles
\begin{equation}
Z_N = \sum_n\exp(-u_0^2\beta p^0_n),
\end{equation}
where $p^0_n$ is the energy of the system in the state labelled by $n$.  
We recall that the grand canonical partition function is given by
\begin{equation}
\Xi = \sum_{N=1}^{\infty}s^NZ_N,
\end{equation}
where $s=e^{\beta\mu}$ is the fugacity and $\mu$ is the chemical
potential.  We point out that the product $\beta\mu$, and thus the fugacity,
is Lorentz invariant (see (3.35), (3.36) and (3.37)).  So the grand canonical
partition function $\Xi$ is Lorentz invariant as well.  As in the case of 
the classical ideal gas the average energy is related to the partition 
function by
\begin{equation}
U = -\frac{\partial}{\partial\beta}\ln \Xi.
\end{equation}
Furthermore, we have also
\begin{equation}
\frac{pV}{kT} = \ln \Xi.
\end{equation}
From (4.4) one can obtain the equation of the state by eliminating
the parameter $s$ with the help of the relation
\begin{equation}
N = s\frac{\partial}{\partial s}\ln \Xi,
\end{equation}
where $N$ is the average number of particles of the gas.  Now taking
into account the statistics of the particles and the form of (4.2)
we obtain the following form of the grand canonical partition function:
\begin{equation}
\Xi = \prod_{\underline{\bm p}}(1\mp s\exp[-u_0^2\beta
p^0(\underline{\bm p})])^{\mp 1},
\end{equation}
where upper and lower sign refer to Bose-Einstein and Fermi-Dirac
gases, respectively, and $p^0(\underline{\bm p})$ is given by (2.13),
(2.14) and (2.15).  Using (4.6) we arrive at the following form of 
the relations (4.3), (4.4) and (4.5):
\begin{eqnarray}
U &=& u_0^2 \sum_{\underline{\bm p}}\frac{p^0(\underline{\bm p})}
{s^{-1}\exp[u_0^2\beta p^0(\underline{\bm p})]\mp 1},\\
\frac{pV}{kT} &=& \mp \sum_{\underline{\bm p}}\ln(1\mp s\exp[-u_0^2
\beta p^0(\underline{\bm p})],\\
N &=& \sum_{\underline{\bm p}}\frac{1}
{s^{-1}\exp[u_0^2\beta p^0(\underline{\bm p})]\mp 1}.
\end{eqnarray}
The above entities are well-defined for $s\ge0$ in the case of
fermions and $0\le s\le1$ for bosons.  In the limit $V\to\infty$
the sums (4.7), (4.8) and (4.9) change to integrals via the
replacement \cite{17}
\begin{equation}
\sum_{\underline{\bm p}}\to \frac{V}{(2\pi)^3}\int d^3\underline{\bm p},
\end{equation}
where we set $\hbar=1$.  Hence, using (2.15), (2.16) and (2.17) we get
\begin{subequations}
\begin{eqnarray}
\left(\frac{U}{V}\right)_b &=& \frac{m^4}{2\pi^2}\int_0^\infty\frac{\cosh^2
t\sinh^2 t}
{s^{-1}\exp(u_0\beta m\cosh t)\mp 1}dt\\
&=& \frac{m^2}{2\pi^2u_0^2\beta^2}\sum_{n=1}^{\infty}\frac{(\pm
1)^{n+1}}{n^2}[3K_2(nu_0\beta m) + nu_0\beta mK_1(nu_0\beta m)]s^n,
\end{eqnarray}
\end{subequations}
\begin{subequations}
\begin{gather}
\left(\frac{U}{V}\right)_l = \frac{u_0}{4\pi^2}\int_{-1}^{1}dx\sqrt{{\bm
u}^2x^2+1}\int_{0}^{\infty}d|\underline{\bm p}|\frac{|\underline{\bm p}|^3}{s^{-1}\exp(u_0\beta\sqrt{{\bm
u}^2x^2+1}|\underline{\bm p}|)\mp 1}\\
= \frac{3}{\pi^2u_0^4\beta^4}\sum_{n=1}^{\infty}\frac{(\pm
1)^{n+1}}{n^4}s^n,
\end{gather}
\end{subequations}
\begin{subequations}
\begin{eqnarray}
&&\left(\frac{U}{V}\right)_t = \frac{m^4}{2\pi^2}\int_0^\infty\frac{\cosh^2
t\sinh^2 t}
{s^{-1}\exp(u_0\beta m\sinh t)\mp 1}dt\\
&&= \frac{m^2}{2\pi^2u_0^2\beta^2}\sum_{n=1}^{\infty}\frac{(\pm
1)^{n+1}}{n^2}[3S_{0,2}(nu_0\beta m) - nu_0\beta mS_{-1,1}(nu_0\beta m)]s^n.
\end{eqnarray}
\end{subequations}
We recall that the indices $b$, $l$, and $t$ refer to bradyons,
luxons and tachyons, respectively.  In the case of the Bose-Einstein gases
we set $s<1$.  The power series expansions
(4.11b) and (4.13b) were obtained with the help of the basic properties
of the Bessel functions and the Lommel functions presented in the
Appendix.  The expansion (4.12b) was derived with the use of the
identity \cite{21}
\begin{equation}
\int_{0}^{\infty}\frac{x^{a-1}}{e^{bx}-c}=\frac{1}{cb^a}\Gamma(a)
\sum_{n=1}^{\infty}\frac{c^n}{n^a},\qquad a>0,
\end{equation}
where $\Gamma(x)$ is the gamma function.  Furthermore, applying the
same technique as with (4.11), (4.12) and (4.13) we find
\begin{subequations}
\begin{eqnarray}
\left(\frac{p}{kT}\right)_b &=& \mp \frac{m^3}{2\pi^2u_0}\int_0^\infty dt\sinh^2 t
\cosh t\ln[1\mp s\exp(-u_0\beta m\cosh t)]\\
&=& \frac{m^2}{2\pi^2u_0^2\beta}\sum_{n=1}^{\infty}\frac{(\pm
1)^{n+1}}{n^2}K_2(nu_0\beta m)s^n,
\end{eqnarray}
\end{subequations}
\begin{subequations}
\begin{gather}
\left(\frac{p}{kT}\right)_l = \frac{u_0\beta}{12\pi^2}\int_{-1}^{1}dx\sqrt{{\bm
u}^2x^2+1}\int_{0}^{\infty}d|\underline{\bm p}|\frac{|\underline{\bm p}|^3}
{s^{-1}\exp(u_0\beta\sqrt{{\bm
u}^2x^2+1}\,|\underline{\bm p}|)\mp 1}\\
= \frac{1}{\pi^2u_0^4\beta^3}\sum_{n=1}^{\infty}\frac{(\pm
1)^{n+1}}{n^4}s^n,
\end{gather}
\end{subequations}
\begin{subequations}
\begin{eqnarray}
\left(\frac{p}{kT}\right)_t &=& \mp \frac{m^3}{2\pi^2u_0}\int_0^\infty dt
\cosh^2 t\sinh t\ln[1\mp s\exp(-u_0\beta m\sinh t)]\\
&=& \frac{m^2}{2\pi^2u_0^2\beta}\sum_{n=1}^{\infty}\frac{(\pm
1)^{n+1}}{n^2}S_{0,2}(nu_0\beta m)s^n.
\end{eqnarray}
\end{subequations}
Finally, we have
\begin{subequations}
\begin{eqnarray}
\left(\frac{N}{V}\right)_b &=& \frac{m^3}{2\pi^2u_0}\int_0^\infty\frac{\sinh^2 t\cosh t}
{s^{-1}\exp(u_0\beta m\cosh t)\mp 1}dt\\
&=& \frac{m^2}{2\pi^2u_0^2\beta}\sum_{n=1}^{\infty}\frac{(\pm
1)^{n+1}}{n}K_2(nu_0\beta m)s^n,
\end{eqnarray}
\end{subequations}
\begin{subequations}
\begin{gather}
\left(\frac{N}{V}\right)_l = \frac{1}{4\pi^2}\int_{-1}^{1}dx
\int_{0}^{\infty}d|\underline{\bm p}|\frac{|\underline{\bm p}|^2}
{s^{-1}\exp(u_0\beta\sqrt{{\bm u}^2x^2+1}\,|\underline{\bm p}|)\mp 1}\\
= \frac{1}{\pi^2u_0^4\beta^3}\sum_{n=1}^{\infty}\frac{(\pm
1)^{n+1}}{n^3}s^n,
\end{gather}
\end{subequations}
\begin{subequations}
\begin{eqnarray}
\left(\frac{N}{V}\right)_t &=& \frac{m^3}{2\pi^2u_0}\int_0^\infty
\frac{\cosh^2 t\sinh^2 t}{s^{-1}\exp(u_0\beta m\sinh t)\mp 1}dt\\
&=& \frac{m^2}{2\pi^2u_0^2\beta}\sum_{n=1}^{\infty}\frac{(\pm
1)^{n+1}}{n}S_{0,2}(nu_0\beta m)s^n.
\end{eqnarray}
\end{subequations}
In the preferred frame, when $u_0=1$, the power series expansions for
bradyons (4.11b), (4.15b) and (4.18b) reduce to the relations
originally obtained by Glaser \cite{22}.  The expressions for tachyons
(4.13b), (4.17b) and (4.20b) in the particular case of the preferred frame
coincide up to a multiplicative normalization constant with the formulas
originally derived by Mr\'owczy\'nski \cite{23}.  We point out that an
immediate consequence of (4.12b) and (4.16b) is the following relation
\begin{equation}
p = \frac{1}{3}\rho,
\end{equation}
where $\rho=U/V$.  Therefore the equation of state for the quantum ideal gas
of massless particles has the same form as in the classical case described by
(3.40) and (3.42).
\subsection{Classical limit}
We now discuss the classical limit when $N/V\to0$ and/or
$T\to\infty$ that is $\beta\to0$.  In view of (4.18a), (4.19a) and (4.20a) it
is clear that in this limit the fugacity $s$ is small as well.
Therefore, we can approximate the series in the formulae (4.11)--(4.20)
by the first terms linear in $s$.  We find
\begin{eqnarray}
\left(\frac{U}{V}\right)_b &=& \frac{m^2}{2\pi^2u_0^2\beta^2}
[3K_2(u_0\beta m) + u_0\beta mK_1(u_0\beta m)]s,\qquad s\ll1\\
\left(\frac{p}{kT}\right)_b &=& \frac{m^2}{2\pi^2u_0^2\beta}
K_2(u_0\beta m)s,\qquad s\ll1\\
\left(\frac{N}{V}\right)_b &=& \frac{m^2}{2\pi^2u_0^2\beta}
K_2(u_0\beta m)s,\qquad s\ll1.
\end{eqnarray}
From (4.23) and (4.24) we obtain the classical equation of state
(see (3.39))
\begin{equation}
pV = kTN, \qquad s\ll1.
\end{equation}
We also have the relation following directly from (4.22) and (4.24)
\begin{equation}
\left(\frac{U}{N}\right)_b = \frac{1}{\beta}\left[3 + u_0\beta m
\frac{K_1(u_0\beta m)}{K_2(u_0\beta m)}\right], \qquad s\ll1,
\end{equation}
which is equivalent to the classical formula (3.10).  Analogously, we
have for luxons
\begin{eqnarray}
\left(\frac{U}{V}\right)_l &=& 
\frac{3}{\pi^2u_0^4\beta^4}s,\qquad s\ll1\\
\left(\frac{p}{kT}\right)_l &=& 
\frac{1}{\pi^2u_0^4\beta^3}s,\qquad s\ll1\\
\left(\frac{N}{V}\right)_l &=& 
\frac{1}{\pi^2u_0^4\beta^3}s,\qquad s\ll1.
\end{eqnarray}
As with bradyons we get from (4.28) and (4.29) the classical
equation of state (4.25).  On the other hand, (4.27) and (4.29) imply
\begin{equation}
\left(\frac{U}{N}\right)_l = \frac{3}{\beta},\qquad s\ll1.
\end{equation}
We have thus obtained the classical expression for the energy of the
luxon ideal gas (3.13).  Finally, consider the case of tachyons.  The
corresponding approximations can be written as
\begin{eqnarray}
\left(\frac{U}{V}\right)_t &=& \frac{m^2}{2\pi^2u_0^2\beta^2}
[3S_{0,2}(u_0\beta m) - u_0\beta mS_{-1,1}(u_0\beta m)]s,\qquad s\ll1\\
\left(\frac{p}{kT}\right)_t &=& \frac{m^2}{2\pi^2u_0^2\beta}
S_{0,2}(u_0\beta m)s,\qquad s\ll1\\
\left(\frac{N}{V}\right)_t &=& \frac{m^2}{2\pi^2u_0^2\beta}
S_{0,2}(u_0\beta m)s,\qquad s\ll1.
\end{eqnarray}
As in the case of bradyons and luxons we obtain from (4.32) and
(4.33) the classical equation of state (4.25).  Furthermore, using
(4.31) and (4.33) we get
\begin{equation}
\left(\frac{U}{N}\right)_t = \frac{1}{\beta}\left[3 - u_0\beta m
\frac{S_{-1,1}(u_0\beta m)}{S_{0,2}(u_0\beta m)}\right], \qquad s\ll1,
\end{equation}
coinciding with the classical expression (3.14).
\subsection{The limit $T\to 0$}
Bearing in mind the technical complexity of calculations in the case
of the degenerate Fermi-Dirac gas as well as the fact that only
bosonic tachyons are admitted as a candidate for a dark matter we
restrict to the Bose gas.  Furthermore, in view of observations of
\cite{3} the limit $T\to 0$ in the case of Bose gas of bradyons is
in fact the non-relativistic limit and will not be discussed herein.
Consider the ideal gas of luxons.  As in the special case of the
photon gas which is well known to be degenerate for all
temperatures, we set $s=1$.  From (4.12), (4.16) and (4.19) we get
\begin{eqnarray}
\frac{U}{V} &=& CT^4,\\
p &=& \frac{1}{3}CT^4,\\
N &=& aT^3V,
\end{eqnarray}
where $C=3\zeta(4)k^4/\pi^2u_0^4=\pi^2k^4/30u_0^4$, and
$a=\zeta(3)k^3/\pi^2u_0^4$;
$\zeta(x)=\sum_{n=1}^{\infty}\frac{1}{n^x}$ is the Riemann zeta
function.  In the case of the photon gas and the preferred frame 
($u_0=1$), the constant $C$ and the constant $a$, taking into 
account the two polarization states of a photon, should be 
multiplied by 2, i.e.\ $C$ is then the Stefan-Boltzmann constant 
and $a=a_{\rm photon}$, respectively.

Now the entropy can be written as (see (3.24))
\begin{equation}
S = k\ln\Xi + \frac{U}{T}.
\end{equation}
The equations (4.38), (4.4), (4.35) and (4.36) taken together yield
\begin{equation}
S=\frac{4}{3}VCT^3.
\end{equation}
As expected, the entropy vanishes at zero temperature.  The formula
(4.39) can be also obtained from the well-known relation
\begin{equation}
S=\int_{0}^{T}\frac{C_V}{T}dT,
\end{equation}
where
\begin{equation}
C_V=\left(\frac{\partial U}{\partial T}\right)_V=4CVT^3,
\end{equation}
following directly from (4.35).

We now study the Bose gas of tachyons in the low-temperature limit.
In this limit one cannot simply replace sums (4.7), (4.8) and (4.9)
by integrals because when $s\simeq 1$ the terms referring to the
ground state can give finite contributions to series.  Therefore, we
separate the sum (4.9) into two contributions --- the number of
particles in the ground state $N_0$ and the number of particles in
the excited states \cite{20}.  Using (4.20b) we find
\begin{equation}
N=N_0+V\frac{m^2}{2\pi^2u_0^2\beta}\sum_{n=1}^{\infty}\frac{
1}{n}S_{0,2}(nu_0\beta m)s^n,
\end{equation}
where
\begin{equation}
N_0=\frac{1}{s^{-1}-1}.
\end{equation}
We recall that the accumulation of bosons in the ground states is
known as Bose-Einstein condensation.  Taking the limit of the strong
degeneration $s=1$ and utilizing for $T\to 0$, that is
$\beta\to\infty$, the asymptotic formula (A.10) we obtain from
(4.42) the relation
\begin{equation}
N = N_0 + V\frac{m}{2\pi^2u_0^3\beta^2}\zeta(2)=N_0+
V\frac{mk^2}{12u_0^3}T^2.
\end{equation}
From (4.44) it follows that the Bose-Einstein condensation occurs at
temperatures lower than the critical one given by
\begin{equation}
T_c = \sqrt{\frac{12Nu_0^3}{Vmk^2}},
\end{equation}
and densities higher than the critical density such that
\begin{equation}
d_c = \frac{mk^2}{12u_0^3}T^2.
\end{equation}
Furthermore, using (4.44) and (4.45) we get
\begin{equation}
\frac{N_0}{N}=\left[1-\left(\frac{T}{T_c}\right)^2\right],\qquad
T\le T_c.
\end{equation}
For temperatures greater than $T_c$ but close enough to $T_c$ to
enable setting $s=1$, we have
\begin{eqnarray}
\frac{U}{V} &=& \frac{m}{\pi^2u_0^3\beta^3}\zeta(3),\\
\frac{p}{kT} &=& \frac{m}{2\pi^2u_0^3\beta^2}\zeta(3),\\
\frac{N}{V} &=& \frac{m}{2\pi^2u_0^3\beta^2}\zeta(2).
\end{eqnarray}
Hence, taking into account (4.49) and (4.50) we arrive at the equation of state 
of the form
\begin{equation}
pV = \frac{\zeta(3)}{\zeta(2)}NkT,\qquad T\gtrsim T_c.
\end{equation}
Using (4.48) and (4.50) we also obtain
\begin{equation}
U = 2\frac{\zeta(3)}{\zeta(2)}NkT,\qquad T\gtrsim T_c.
\end{equation}
Therefore the specific heat is 
\begin{equation}
C_V = 2\frac{\zeta(3)}{\zeta(2)}Nk,\qquad T\gtrsim T_c.
\end{equation}
For temperatures lower than $T_c$ one should use $N-N_0$ instead of
$N$.  Therefore the equation of state takes the form
\begin{equation}
p = \frac{m}{2\pi^2u_0^3}\zeta(3)k^3T^3,\qquad T\le T_c,
\end{equation}
and (4.52) is replaced by
\begin{equation}
U = \frac{m}{\pi^2u_0^3}\zeta(3)k^3T^3V,\qquad T\le T_c.
\end{equation}
Using (4.38), (4.4), (4.54) and (4.55) we obtain the following
formula on the entropy:
\begin{equation}
S = \frac{3m}{2\pi^2u_0^3}V\zeta(3)k^3T^2,\qquad T\le T_c,
\end{equation}
implying vanishing of the entropy at zero temperature.  The relation
(4.56) is also a consequence of (4.40) and the expression on the
specific heat such that
\begin{equation}
C_V = \frac{3m}{\pi^2u_0^3}V\zeta(3)k^3T^2,\qquad T\le T_c,
\end{equation}
following immediately from (4.55).  The discussion of the relations 
satisfied by the thermodynamic functions for the non-relativistic Bose 
gas in the case of $T\gtrsim T_c$, and $T\le T_c$ can be found in the book
\cite{24}. 
\section{Conclusions}
In this work we have derived the Lorentz covariant form of thermodynamic
functions for the relativistic ideal gas in both classical and quantum
cases.  We stress that the applied
approach based on the concept of the preferred frame is the only one 
which enables formulation of covariant statistical mechanics of tachyons.
On the other hand, an advantage of the formalism introduced in this paper
is that it allows to study from the unique point of view all kinds of
relativistic gases including bradyon, luxon, and tachyon ones.  Bearing in
mind the possible applications of the observations of this work we point
out the specific heat of the tachyon gas showing behavior analogous to
Schottky anomaly (see Fig.\ 3), and the formula (4.43) on the critical
temperature for the Bose condensation which seem to be of importance for
the determination of the mass of a tachyon.  The results obtained in this 
paper can be also helpful in discussion of the dark matter and dark energy 
in astrophysics and cosmology.  Indeed, the observations of stars motion 
in galactics, galactic clusters, cluster masses as inferred from 
gravitational lensing strongly suggest existence of an exotic dark matter 
component of the universe.  Moreover, from observations of supernova 
IA populations we know that the universe accelerates which demands existence 
of exotic dark energy.  A possible attempts to explanation of the exotic 
content of the universe need some radical extension of standard physics.  
One of candidates is tachyonic fluid (see for example \cite{26}) 
or the rolling tachyon field \cite{9,27}.
\section*{Acknowledgements}
This paper has been supported by University of Lodz grant.  We would like
to thank Janusz J{\c e}drzejewski for helpful comments.
\appendix*
\section{}
We first recall some properties of the modified Bessel functions
(Macdonald functions) $K_\nu(x)$.  These functions have the integral
representation of the form
\begin{equation}
K_\nu(x) = \int_{0}^{\infty}e^{-x\cosh t}\cosh\nu t\,dt,\qquad x>0.
\end{equation}
They satisfy the following recurrence relations:
\begin{equation}
K'_\nu(x) =\frac{\nu}{x}K_\nu(x)-K_{\nu+1}(x)
=-\frac{\nu}{x}K_\nu(x)-K_{\nu-1}(x),
\end{equation}
where prime designates differentiation.  We have the asymptotic
formulas
\begin{eqnarray}
K_n(x) &=& \frac{1}{2}(n-1)!\left(\frac{2}{x}\right)^n,\qquad x\ll 1,\\
K_\nu(x) &=& \sqrt{\frac{\pi}{2x}}e^{-x},\qquad x\gg 1.
\end{eqnarray}

We now briefly sketch the basic properties of the Lommel functions
$S_{\mu,\nu}(x)$ \cite{21,25}.  The Lommel functions $S_{0,\nu}(x)$ are
given by
\begin{equation}
S_{0,\nu}(x) = \int_{0}^{\infty}e^{-x\sinh t}\cosh\nu t\,dt
= \frac{x}{\nu}\int_{0}^{\infty}e^{-x\sinh t}\cosh t\sinh\nu
t\,dt,\qquad x>0.
\end{equation}
The Lommel functions $S_{\nu,\nu}(x)$ can be expressed in terms of
the Struve functions ${\bf H}_\nu(x)$ and the Bessel functions
$Y_\nu(x)$ (Neumann functions) also designated by $N_\nu(x)$.  We have
\begin{equation}
S_{\nu,\nu}(x) = 2^{\nu-1}\sqrt{\pi}\Gamma(\nu + \hbox{$\scriptstyle
1\over 2$})[{\bf H}_\nu(x)-Y_\nu(x)].
\end{equation}
The recurrence relations for $S_{\mu,\nu}(x)$ are of the form
\begin{gather}
S_{\mu+2,\nu}(x) = x^{\mu+1}-[(\mu+1)^2-\nu^2]S_{\mu,\nu}(x),\\
S'_{\mu,\nu}(x) = \frac{\nu}{x}S_{\mu,\nu}(x)+(\mu-\nu-1)S_{\mu
-1,\nu +1}(x) = -\frac{\nu}{x}S_{\mu,\nu}(x)+(\mu+\nu-1)S_{\mu
-1,\nu -1}(x).
\end{gather}
For small and large argument $S_{0,2}(x)$ can be approximated as
\begin{eqnarray}
S_{0,2}(x) &=& \frac{2}{x^2},\qquad x\ll 1,\\
S_{0,2}(x) &=& \frac{1}{x},\qquad x\gg 1.
\end{eqnarray}
We remark that $S_{0,2}(x)$ and $K_2(x)$ are approximated by the
same function in the limit $x\ll 1$.


\begin{references}
\bibitem{1}F. J\"uttner, Ann. Phys. {\bf 34}, 856 (1911); {\bf 35}, 145 (1911).
\bibitem{2}D. Ter Haar, H. Wergeland, Phys. Rep. {\bf 1}, 31 (1971).
\bibitem{3}C. Arag\~ao de Carvalho and S. Goulart Rosa Jr, J. Phys. A 
{\bf 13}, 3233 (1980).
\bibitem{4}J. Rembieli\'nski, K.A. Smoli\'nski and G. Duniec,
Found. Phys. Lett. {\bf 14}, 487 (2001).
\bibitem{5}J. Rembieli\'nski, Internat. J. Modern Phys. A {\bf 12}, 1677 (1997).
\bibitem{6}G. Feinberg, Phys. Rev. {\bf 159}, 1089 (1967).
\bibitem{7}P.C.W. Davies, Internat. J. Theoret. Phys. {\bf 43}, 141 (2004);
A. Das {\em et al}, Phys. Rev. D {\bf 72}, 043528 (2005).
\bibitem{8}J.S. Bagla, H.K. Jassal and T. Padmanabhan, Phys. Rev. D {\bf 67},
063504 (2003); E.J. Copeland {\em et al}, Phys. Rev. D {\bf 71}, 043003 
(2005); A. Das {\em et al}, Phys. Rev. D {\bf 72}, 043528 (2005);
E. Calcagni and A.R. Liddle, Phys. Rev. D {\bf 74}, 043528 (2006).
\bibitem{9}S. Mukohyama, Phys. Rev. D {\bf 66}, 024009 (2002).
\bibitem{10}F.A.E. Pirani, Phys. Rev. D {\bf 1}, 3224 (1970); J.K. Kowalczy\'nski, 
Internat. J. Theoret. Phys. {\bf 23}, 27 (1984).
\bibitem{11}E.C.G. Sudarshan, in: E. Recami ed., {\em Tachyons, Monopoles and
Related Topics} (North-Holland, New York, 1978), pp. 43--46.
\bibitem{12}K. Kamoi and S. Kamefuchi, in: E. Recami ed., {\em Tachyons, Monopoles 
and Related Topics} (North-Holland, New York, 1978), pp. 159--167.
\bibitem{13}J. Rembieli\'nski, Phys. Lett. A {\bf 78}, 33 (1980).
\bibitem{14}P. Caban and J. Rembieli\'nski, Phys. Rev. A {\bf 59}, 4187 (1999).
\bibitem{15}G. Amelino-Camelia, Internat. J. Modern Phys. D {\bf 11}, 35 (2002);
J. Magueijo and L. Smolin, Phys. Rev. D {\bf 67}, 044017 (2003).
\bibitem{16}T. Jacobson and D. Mattingly, Phys. Rev. D {\bf 64}, 024028 (2001).
\bibitem{17}D. Colladay and V.A. Kosteleck\'y, Phys. Rev. D {\bf 55}, 6760 (1997);
{\em ibid\/} Phys. Rev. D {\bf 58}, 116002 (1998); V.A. Kosteleck\'y, Phys. Rev. D 
{\bf 69}, 105009 (2004).
\bibitem{18}S. Mr\'owczy\'nski, Lett. Nuovo Cim. {\bf 38}, 247 (1983).
\bibitem{19}H.B. Callen, {\em Thermodynamics and an Introduction to
Thermostatics} (Wiley, New York, 1985).
\bibitem{20}K. Huang, {\em Statistical Mechanics} (Wiley, New York, 1987).
\bibitem{21}I.S. Gradshteyn and I.M. Ryzhik, {\em Tables of Integrals, 
Series, and Products} (Academic Press, New York, 2000).
\bibitem{22}W. Glaser, Z. Phys. {\bf 94}, 677 (1935).
\bibitem{23}S. Mr\'owczy\'nski, Nuovo Cim. B {\bf 81}, 179 (1984).
\bibitem{24}J. {\L}opusza\'nski and A. Pawlikowski, {\em Statistical Physics} (PWN, 
Warsaw, 1969).
\bibitem{25}H. Bateman, {\em Higher Transcendental Functions}, vol. 2,
(McGraw-Hill, New York, 1953).
\bibitem{26}P.C.W. Davies, Internat. J. Theoret. Phys. {\bf 43}, 141 (2004).
\bibitem{27}G. Gibbons, Classical Quantum Gravity {\bf 20}, 5231 (2003).
\end{references}
\end{document}